%% file: ase-submit.tex
\documentclass[sigconf]{acmart}

\usepackage[switch]{lineno}
\newcommand{\toolname}{\textsc{PSearch}\xspace}

\usepackage{tcolorbox}
\tcbuselibrary{breakable}
\usepackage{amsmath,amssymb,amsfonts}
\usepackage{pifont}
\usepackage{enumerate}
\usepackage{ragged2e}
\usepackage{makecell}
\usepackage{stfloats}
\usepackage[vlined, ruled]{algorithm2e}
\usepackage{algorithmic}
\usepackage{graphicx}
\usepackage{textcomp}
\usepackage{multirow}
\usepackage{subfigure}
\usepackage{diagbox}
\usepackage{listings}
\usepackage{float}

\usepackage{url}
\usepackage{hyperref}

\definecolor{deepblue}{rgb}{0,0,0.5}
\definecolor{deepgreen}{rgb}{0,0.5,0}
\definecolor{deepred}{rgb}{0.6,0,0}
\definecolor{darkorange}{RGB}{255,140,0}
\definecolor{lightgray}{rgb}{0.93,0.93,0.93}
\definecolor{deepgray}{rgb}{0.25,0.25,0.25}

\hypersetup{
    colorlinks=true,
    linkcolor=blue,
    filecolor=blue,      
    urlcolor=blue,
    citecolor=blue,
}

\lstdefinelanguage{Java}{
	basicstyle=\small\ttfamily,
	numberstyle=\color{deepgray},
	stepnumber=1,
	numbersep=8pt,
	showstringspaces=false,
	breaklines=true,
	frame=lines,
	backgroundcolor=\color{lightgray},
	commentstyle=\color{deepgreen},
	keywordstyle=\color{deepblue},
	stringstyle=\color{deepred},
	tabsize=4,
	captionpos=b,
	morekeywords={public, class, void, int, if, else, for, while, return, true, false},
	emph={String, System},
	emphstyle=\color{darkorange},
	alsoletter={.,;:[]()},
}

\usepackage{multirow}
\usepackage[utf8]{inputenc}
\usepackage{graphicx}
\usepackage{textcomp}
\usepackage{color,xcolor}
\usepackage{url}
\usepackage{graphicx}
\usepackage{subfigure}
\usepackage{stmaryrd}
\usepackage{verbatimbox}
\usepackage{enumerate}
\usepackage[shortlabels]{enumitem}
\usepackage{bm}

\usepackage[justification=centering]{caption}
\newtheorem{definition}{Definition}
\usepackage{makecell}
\usepackage{booktabs}

\newcommand{\find}[2]{
\begin{tcolorbox}[toprule=0mm,bottomrule=0mm,left=1pt,right=2pt,top=2pt,bottom=2pt,breakable]%
\em #1
\end{tcolorbox}
}

\usepackage{multirow}

\usepackage{enumitem}
\usepackage{colortbl}

\newcommand{\finding}[2]{
\begin{center}
\begin{tcolorbox}[leftrule=0mm,toprule=0mm,bottomrule=0mm,rightrule=0mm,left=1pt,right=2pt,top=0pt,bottom=0pt,breakable]
\textbf{Answer to RQ{#1}:}
{#2}
\end{tcolorbox}
\end{center}
}

\usepackage[normalem]{ulem}
\newcommand{\revise}[1]{{\color{black}{#1}}}
\newcommand{\delete}[1]{\iffalse{#1}\fi}

\AtBeginDocument{%
  \providecommand\BibTeX{{%
    \normalfont B\kern-0.5em{\scshape i\kern-0.25em b}\kern-0.8em\TeX}}}
    
\begin{document}

\title{\toolname{}: Search-based Patch Generation in the Era of LLM-based Automated Program Repair}

\author{Haichuan Hu}
\email{huhaichuan2024@gmail.com}
\orcid{0009-0002-3007-488X}
\affiliation{
\institution{Nanjing University of Science and Technology}
\city{Nanjing}\country{China}
}

\author{Ye Shang}
\email{yeshang@smail.nju.edu.cn}
\orcid{0009-0000-8699-8075}
\affiliation{
\institution{Nanjing University}
\city{Nanjing}\country{China}
}

\author{Weifeng Sun}
\email{weifeng.sun@cqu.edu.cn}
\orcid{0000-0001-6013-1369}
\affiliation{
\institution{Singapore Management University}
\country{Singapore}
}

\author{Quanjun Zhang}
\authornote{corresponding author.}
\email{quanjunzhang@njust.edu.cn}
\orcid{0000-0002-2495-3805}
\affiliation{
\institution{Nanjing University of Science and Technology}
\city{Nanjing}\country{China}
}

\begin{abstract}
Large Language Models (LLMs) have substantially advanced Automated Program Repair (APR), yet most existing LLM-based APR methods still rely on trial-and-error to generate patches. Such a strategy explores candidate patches in a weakly structured manner, making it difficult to assess the future potential of search directions and allocate search budget effectively.
To address this limitation, we propose \toolname{}, a search-based patch generation framework for LLM-based APR centered on iterative patch evaluation and refinement. Instead of treating patch generation as repeated independent sampling, \toolname{} maintains a structured search state over intermediate patches, continuously evaluates the promise of explored search paths, and prioritizes the most promising ones for further refinement. This design enables \toolname{} to abandon weak directions early and progressively approach correct fixes through long-horizon search. Importantly, \toolname{} can be integrated with different search algorithms, while our current implementation adopts Monte Carlo Tree Search as one effective instantiation.
We evaluate \toolname{} on five widely used bug and vulnerability benchmarks. Experimental results show that \toolname{} correctly repairs 201 out of 835 bugs in Defects4J, outperforming all 12 state-of-the-art baselines. \toolname{} also fixes 27 of 79 vulnerabilities in VUL4J and resolves 164 of 300 issues in SWE-Bench-Lite. Moreover, with a patch size of 16, \toolname{} reduces monetary cost to roughly 50\% of strong baselines while maintaining superior repair effectiveness. These results highlight the effectiveness of \toolname{} for improving LLM-based APR.
\end{abstract}



\maketitle

\section{Introduction}

Automated Program Repair (APR)~\cite{zhang2023survey} aims to automatically locate software defects and generate fixes that restore program correctness.
A typical APR workflow involves generating candidate patches and validating them against available tests, followed by manual inspection to determine whether the patch is genuinely correct.
Traditional APR techniques can be generally classified into three categories: template-based~\cite{44,45}, heuristic-based~\cite{Simfix:2018,10.1145/3468264.3468600}, and constraint-based~\cite{196,106} methods.
Among them, template-based APR has long been regarded as one of the most effective paradigms, but its dependence on predefined transformation templates limits its ability to handle previously unseen bugs.

To overcome the limited coverage of handcrafted repair patterns, researchers have introduced a large body of learning-based APR techniques, which learn bug-fixing knowledge from existing code repositories and bug-fix pairs~\cite{yuan2022circle}.
Compared with traditional APR, learning-based APR offers stronger generalization and can repair bugs beyond manually designed templates.
Recently, with the rapid progress of Large Language Models (LLMs) in software engineering tasks~\cite{zhang2023surveyse,zhang2025large,shang2025large,zhang2025exploring}, LLM-based APR has emerged as a highly promising direction~\cite{zhang2024survey}.
Hossain et al.~\cite{DBLP:journals/pacmse/Hossain0Z0CLNT24} provide a comprehensive analysis of how prompts and contexts affect the effectiveness of LLM-based APR.
ChatRepair~\cite{DBLP:conf/issta/Xia024}, for example, uses GPT-3.5 to repair 162 bugs on Defects4J~\cite{DBLP:conf/issta/JustJE14}, representing one of the most influential prompting-based APR systems.
Other recent studies~\cite{DBLP:conf/icse/XiaWZ23,zhang2024pre,li2025hybrid,xu2025aligning} further demonstrate that LLMs can be effective in a variety of repair settings.

Despite this progress, the dominant patch generation paradigm in current LLM-based APR remains largely \emph{sample-and-validate}.
A model generates one or more candidate patches, the system executes tests, and subsequent attempts are produced either independently or with limited feedback from previous failures.
Although simple and practical, this strategy still treats patch search as a series of loosely connected local trials, which leads to two major limitations.
First, \textbf{existing methods often fail to estimate the \emph{future potential} of a patch search direction.}
A currently incorrect patch may still be a highly promising intermediate state if it partially captures the bug cause and can be refined into a correct patch.
Conversely, some candidates may look superficially plausible or pass a few tests but actually steer the search toward dead ends.
Without an explicit mechanism for evaluating the promise of intermediate search states and the trajectories they induce, current systems can easily over-invest in low-value directions or prematurely discard promising ones.
Second, \textbf{current methods generate patches inefficiently under limited budget.}
Because generated patches are frequently explored in a weakly structured or memoryless way, the system may repeatedly revisit similar mistakes, continue repairing an unproductive path for too long, or fail to reuse useful information gathered from previous attempts.
This inefficiency becomes especially severe for complex bugs, where the correct fix often cannot be found in a single step and instead requires multiple rounds of informed refinement.

To address the two limitations, we propose \toolname{}, a search-based patch generation framework for LLM-based APR that organizes patch generation as iterative exploration over intermediate search states.
\toolname{} maintains a structured search space of explored patches, repeatedly selects a promising partial patch, refines it with LLM reasoning and self-reflection, evaluates the resulting candidates using adaptive reward signals, and updates the global search state to guide subsequent iterations.
This design allows the framework to backtrack from weak directions, preserve high-potential partial solutions, and progressively approach correct patches through long-horizon refinement.
Importantly, \toolname{} supports flexible search-based patch generation: the search controller is modular and can be instantiated with different search algorithms.
In this paper, we implement the controller using Monte Carlo Tree Search (MCTS), because it provides an effective balance between exploration and exploitation.
Compared with prior LLM-based patch generation techniques, \toolname{} has the following advantages.
(1) \textbf{State-aware patch search.}
\toolname{} explicitly models intermediate search states and evaluates their potential, enabling the system to distinguish promising search directions from unproductive ones.
This helps allocate repair budget more rationally than independent patch sampling and makes the framework less prone to being trapped in locally appealing but globally unproductive trajectories. 
(2) \textbf{Long-horizon iterative refinement.}
Instead of treating each generation attempt as a fresh restart, \toolname{} supports multi-step refinement over partial patches.
This is especially useful for complex bugs whose fixes emerge only after several rounds of correction and adjustment.
(3) \textbf{Flexibility and generality.}
\toolname{} is compatible with different backbone LLMs and is not inherently coupled with a single search algorithm.

We evaluate \toolname{} on five widely used benchmarks spanning general bug repair, issue repair, and vulnerability repair. On Defects4J, \toolname{} correctly fixes 201 out of 835 bugs, outperforming all 12 state-of-the-art baselines. It also repairs 27 out of 79 vulnerabilities in VUL4J and resolves 164 out of 300 issues in SWE-Bench-Lite, while reducing monetary cost by more than 50\% at a patch size of 16.

In summary, this paper makes the following contributions:

\begin{itemize}[leftmargin=*]
    \item We propose \toolname{}, a search-based patch generation framework for LLM-based APR. The framework is designed to improve patch search by evaluating the potential of intermediate patches and iteratively refining the most promising ones.

    \item We instantiate \toolname{} with MCTS and adaptive patch evaluation, and show that this implementation substantially improves repair effectiveness over 12 state-of-the-art baselines and multiple strong backbone LLMs.

    \item We demonstrate that the benefits of \toolname{} extend beyond standard APR benchmarks. The framework generalizes across multiple languages, bug types, repository-level issue repair, and vulnerability repair.
\end{itemize}

\section{Background and Motivation}

\subsection{Automated Program Repair}
Automated Program Repair (APR) aims to automatically localize and fix program bugs. Traditional APR techniques can generally be classified into heuristic-based~\cite{Simfix:2018,10.1145/3468264.3468600}, constraint-based~\cite{196,106}, and template-based~\cite{44,45} methods. Although these methods have achieved promising results, their repair capability is often limited by handcrafted rules, search heuristics, or predefined transformation patterns.
To improve repair generalization, recent APR research has increasingly explored learning-based methods. These methods learn bug-fixing knowledge from large-scale code corpora or bug-fix pairs, and thus can repair bugs beyond manually designed templates~\cite{VulRepair,49,TransR}. Among them, Neural Machine Translation (NMT)-based approaches have been extensively studied in recent years~\cite{DBLP:conf/icse/MengWZSLH23,DBLP:conf/icse/ZhuSZXZ23,DBLP:conf/kbse/YeML0M22,DBLP:conf/icse/YeMM22,DBLP:conf/sigsoft/ZhuSXZY0Z21,DBLP:conf/icse/JiangL021}. These methods view APR as a translation task that transforms buggy code into correct code.
More recently, Large Language Models (LLMs) have become a major technique for APR due to their strong code understanding and generation capabilities~\cite{DBLP:conf/sigsoft/XiaZ22,DBLP:conf/iclr/FriedAL0WSZYZL23,DBLP:conf/kbse/XiaDZ23,hu2025repair,hu2024gpto1killbugsevaluation}. Compared with earlier learning-based methods, LLM-based APR reduces the dependence on high-quality training datasets and enables zero-shot or few-shot repair. Xia et al.~\cite{DBLP:conf/icse/XiaWZ23} conducted an extensive study of LLM-based APR using models such as Codex~\cite{DBLP:journals/corr/abs-2107-03374}, GPT-NeoX~\cite{DBLP:journals/corr/abs-2204-06745}, CodeT5~\cite{DBLP:conf/emnlp/0034WJH21}, and InCoder~\cite{DBLP:conf/iclr/FriedAL0WSZYZL23}, showing the strong potential of LLMs for APR. Subsequent approaches further improved repair effectiveness through better prompting, richer contexts, and iterative interaction. For example, ChatRepair~\cite{DBLP:conf/issta/Xia024} leverages GPT-3.5 to achieve state-of-the-art performance on Defects4J.
Despite this progress, effective \emph{patch search} remains a central bottleneck in APR, especially for LLM-based methods. In practice, a repair system must do more than generate good patches in isolation: it must decide which intermediate candidates deserve further refinement, which directions should be abandoned, and how limited search budget should be allocated across competing search paths. Therefore, the key challenge is not only to improve patch generation quality, but also to enable search-based patch generation that can effectively evaluate intermediate search states and prioritize the most promising search trajectories.

\subsection{Motivation Example}
To better illustrate the limitations of existing LLM-based patch generation methods, we present a motivation example in this section.
We use a real-world bug, Jsoup\_54 from Defects4J, and evaluate three representative generation settings on it: direct sampling, single-path iterative generation, and genetic search.
We find that none of the three settings works effectively under a limited budget.
Since the order of function-call parameters is incorrect and there are many possible values for those parameters, it is difficult to find the correct solution through direct sampling within a limited sample size. Specifically, the single-path search strategy keeps refining the first incorrect direction it chooses and ignores other promising alternatives. The genetic algorithm maintains multiple candidates, but without a sufficiently reliable path-evaluation mechanism, it still fails to preserve the candidates that are most useful for future refinement. To address this issue, we allow the model to evaluate the quality of explored patches during patch generation.
Patches with lower quality are discarded in time, while patches with higher quality are preserved and further refined in subsequent search.
This design helps the patch generation process avoid wasting budget on low-value directions and instead focus on more promising candidates. After eight attempts, the model successfully fixes Jsoup\_54.
The key insight from this example is that effective patch generation requires not merely producing more patches, but continuously assessing patch quality and prioritizing higher-potential paths.

\section{Approach}

\begin{figure*}[htb]
  \centering
  \includegraphics[width=0.95\linewidth]{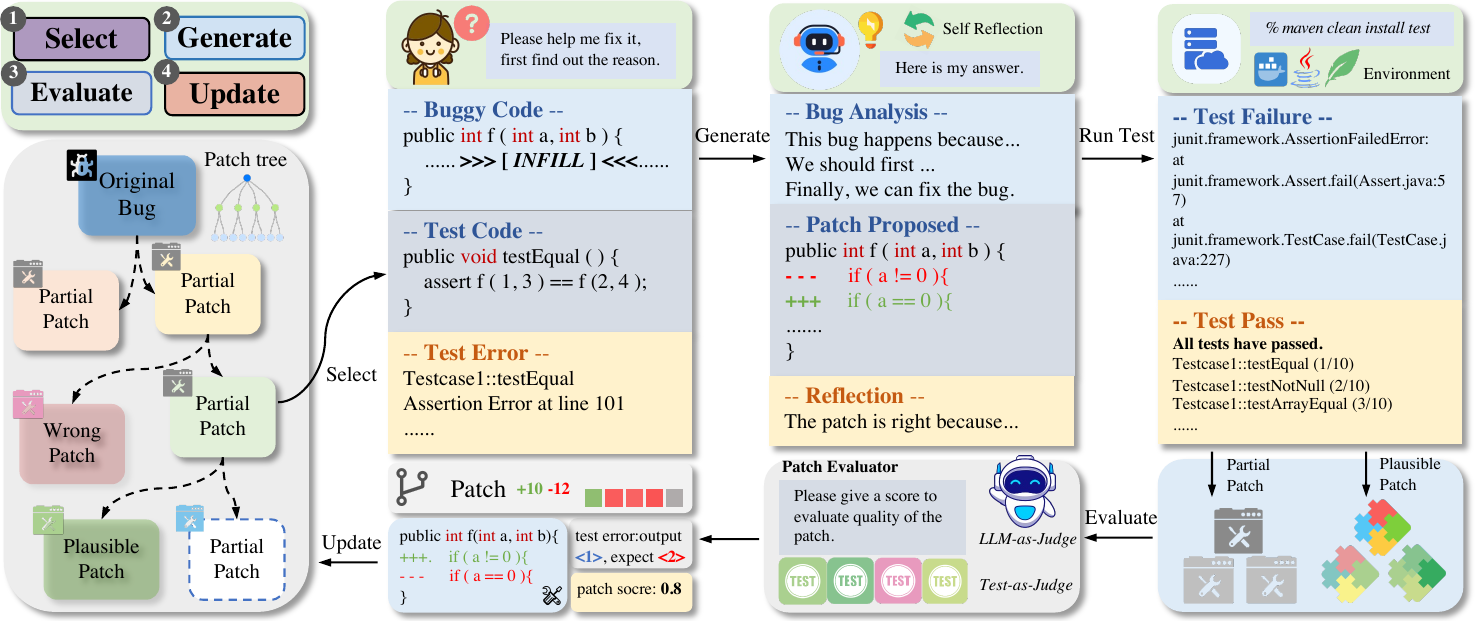}
  \caption{Overview of \toolname{}} 
  \label{overview}
\end{figure*}

In this section, we introduce the concepts used in \toolname{}, the task formulation of \toolname{}, the overall workflow of \toolname{}, and each stage within the process.
Figure~\ref{overview} illustrates the workflow of \toolname{}, which consists of four iterative stages.
First, the framework selects a promising intermediate patch as the next refinement target.
Second, it generates new candidate patches conditioned on that intermediate state.
Third, it evaluates the generated candidates to estimate both correctness and future search potential.
Finally, it updates the global search state so that later iterations can make more informed decisions.
This design reflects the central idea of \toolname{}: patch generation should be organized as iterative search over partial solutions, rather than independent patch sampling.
The search controller itself is modular.
In the current implementation, we instantiate it with a tree-structured controller based on MCTS, but the framework is not intrinsically tied to that choice.

\subsection{Concepts}
Before introduction, first we provide explanations of the concepts used in \toolname{}.

\begin{itemize}[leftmargin=*]
    \item \textbf{Patch Search State.} \toolname{} maintains a structured search state over explored patches. In our current implementation, this state is represented as a patch tree. The root node of the tree is the original bug, which can be considered a special patch.
    \item \textbf{Parent Patch.} If patch $a$ is the parent patch of patch $b$, it means we generate $b$ by refining $a$.
    \item \textbf{Child Patch.} If patch $b$ is a child patch of patch $a$, it means $b$ represents a follow-up repair attempt derived from $a$.
    \item \textbf{Patch Size.} Number of candidate patches applied to a bug.
\end{itemize}

\subsection{Task Formulation}

Suppose $\mathcal{D} = ({P_i^{*}, B_i, T_i)}^{|\mathcal{D}|}_{i=1}$ is a defect dataset consisting of $|\mathcal{D}|$ bugs, each bug $B_i$ paired with a test suite $T_i = [t_1, \ldots, t_m]$ and a developer patch $P_i^{*}$.
For bug $B_i$, the patch search task involves searching for an N-size patch set $P_i = \{{p_j}\}^{N}_{j=1}$ based on $B_i$ and $T_i$, where at least one candidate patch \(p_j \in P_i\) is semantically equivalent to \(P_i^{*}\).
The patch search task is defined as follows.

\find{
\begin{definition}
\label{def:patch_search_task}
\textbf{Patch Search Task:}\\
Given a bug $B_i$ with $m$ test cases $T_i=\left[t_1, \ldots, t_m\right]$ and a patch set output $P_i=\left[p_1, \ldots, p_N\right]$, 
the problem of patch search is formalized as:
\begin{displaymath}
\small
P_{\theta}(P_i|B_i,T_i)=\prod_{j=1}^{N}P_{\theta}
(p_j|p_1, \cdots, p_{j-1}; t_1, \cdots, t_m; B_i) 
\end{displaymath}
\end{definition}
}

\toolname{} further adopts a state-aware strategy to improve traditional patch search.
In each iteration, \toolname{} takes the search state $\tau_{i-1}$ from the previous iteration as input, and outputs a newly generated patch $p_i$ together with the updated search state $\tau_i$.
Based on Definition~\ref{def:patch_search_task}, the state-aware patch search task is defined as follows.

\find{
\begin{definition}
\textbf{State-Aware Patch Search Task:}\\
State-aware patch search consists of two steps.
First, generate a new patch $p_i$ based on the former search state $\tau_{i-1}$, and then update the search state $\tau_{i-1}$ with $p_i$ to output $\tau_{i}$. $\oplus$ represents the update operation.
Thus, given a bug $B_i$, test cases $T_i$, and the current search state $\tau_{i-1}$, the problem is formalized as:
\begin{displaymath}
\small
P_{\theta}(P_i|B_i,T_i)=\prod_{j=1}^{N}P_{\theta}
(p_j|\tau_{i-1}; B_i; t_1, \cdots, t_m) 
\end{displaymath}
\begin{displaymath}
\small
\tau_i=\tau_{i-1}\oplus p_i
\end{displaymath}
\end{definition}
}

\subsection{Stages \& Modules} \label{stage-module}
Given a buggy program, the repair process begins by treating the original buggy code as a special form of patch, which is initialized as the root state of the search structure.
As the repair proceeds, newly generated patches are incrementally added as refinements of previously explored patches.

\subsubsection{Patch Selection} \label{stage-module-1}
In the patch selection stage, \toolname{} aims to identify the most promising intermediate patch from the current search state, which will then be refined into new candidate patches in subsequent stages.
The essential requirement of this module is to balance \emph{exploitation} of patches that have already shown strong potential with \emph{exploration} of less-visited patches that may still lead to better repair outcomes.

The framework itself does not require a single fixed selection strategy.
In this work, we instantiate the selector with the Upper Confidence Bound for Trees (UCT), which is a common criterion in MCTS.
UCT takes into account both the average quality of child patches and the degree of exploration, thus providing a more comprehensive assessment of a patch's future potential.
A higher UCT indicates that continuing search from the corresponding patch is more likely to lead to a plausible repair.
In our implementation, UCT is defined as follows:

\begin{equation}
 UCT_j = \bar{X}_j + C  \sqrt{\frac{2 \ln N_C}{N_j}} .
 \label{eq:1}
\end{equation}
Where \( \bar{X}_j \) is the average reward of child node $j$, \( N_C \) is the total number of visits to the parent node, and \( N_j \) is the number of times that child node $j$ has been visited.
$C$ is a constant for balancing exploitation and exploration.
During the patch selection stage, \toolname{} calculates the UCT value for each candidate state and selects the patch with the highest UCT from the existing search structure.

\subsubsection{Patch Generation} \label{stage-module-2}
In the patch generation stage, \toolname{} aims to generate new candidate patches based on the partial patch returned by the patch selection stage.
Instead of generating patches from scratch at every attempt, the framework conditions generation on the current patch search state so that search can proceed incrementally.

To this end, \toolname{} employs a self-refinement strategy that integrates APR-specific reasoning and self-reflection, thereby enhancing the quality of the model's outputs.
Specifically, \toolname{} interprets the current state of the bug from the selected partial patch and performs a comprehensive analysis of the buggy lines and the errors reported by the test cases.
Based on this analysis, it modifies and refines the partial patch to generate new candidate patches.
For a given LLM $\pi$, the conditional probability distribution of generating a new patch $a^{\prime}$ from a previously explored partial patch $a$ is formalized as follows:

\begin{equation}
\pi(a^{\prime}|a) = \prod_{k=1}^{K} \pi(a^{\prime}_k|a^{\prime}_{<k}, a).
 \label{eq:2}
\end{equation}
Where $k$ represents the $k$-th token of $a^{\prime}$.

\textbf{APR-specific reasoning.}
Before generating the new patch $a^{\prime}$, \toolname{} prompts the model to explicitly analyze the bug, the currently selected patch, and the observed test failures.
The purpose of this step is to make each refinement attempt more informed and interpretable.
By articulating what is already correct, what remains wrong, and which repair action should be applied next, the model can better produce patches that represent meaningful progress rather than unrelated resampling.

\textbf{Self-Reflection.}
After generating a patch $a^{\prime}$, we further prompt the model to reflect on its output through a self-reflection mechanism.
This process encourages the model to critically evaluate the generated patch, identify potential errors, and revise its solution accordingly.
By enabling this self-correction step, the model is able to produce higher-quality and more reliable patches.

\subsubsection{Patch Evaluation} \label{stage-module-3}

In the patch evaluation stage, \toolname{} aims to estimate both the correctness and the future potential of the patches returned in the previous stage, thus guiding the framework toward high-value search directions.
After the Patch Generation stage, we execute test cases to verify the correctness of the generated patches $a^{\prime}$.
If a patch passes all test cases, it is marked as a plausible patch and retained for further human inspection.
If it fails any test case, it is not immediately discarded; instead, it is treated as an intermediate patch that may still be useful for future refinement and is added as a new node to the search structure.
Following this, \toolname{} performs a quality assessment of the generated patches using two evaluation strategies: LLM-as-Judge and Test-as-Judge.

\textbf{LLM-as-Judge}.
This strategy utilizes LLMs to score the quality of generated patches in scenarios where test coverage is limited.
For example, a significant portion of bugs in the Defects4J dataset are associated with only a single fault-triggering test case.
In such cases, relying solely on test outcomes may lead to sparse reward signals, which reduces the accuracy of evaluation and the effectiveness of repair.
To address this issue, \toolname{} employs LLM-as-Judge to evaluate patch quality based on semantic and contextual information rather than exclusively on test results.
The input to the evaluation model includes test cases, test results, buggy code, candidate patches, surrounding code context, the reasoning trace, and the reflection output.
The raw score generated by the LLM is further normalized under defined constraints to ensure consistency and fairness in reward computation.
The reward $R(a)$ is defined as:

\begin{equation} \label{reward_equation}
R(a) =
\begin{cases}
-1, & \text{if the patch fails to compile} \\
-0.5, & \text{if the patch repeats previous errors} \\
0, & \text{if } Score(a) \leq 0 \\
1, & \text{if } Score(a) \geq 100 \\
\frac{Score(a)}{100}, & \text{otherwise}
\end{cases}.
\end{equation}

\begin{equation} \label{Qvalue}
\mathbb{E}[R] = \frac{1}{T} \sum_{i=1}^{T} R_i
\end{equation}

Since the scores provided by the LLM may fluctuate, we further estimate the expected reward of a patch.
As shown in Equation~\ref{Qvalue}, the expected value of $R$ is obtained by sampling the reward $T$ times (set to 5 in our study) and taking the average, which helps stabilize evaluation and balance occasional noise in LLM judgments.
The resulting patch $a^{\prime}$ is then encapsulated into a tree node and added to the search structure.
In addition, we adopt a self-evaluation strategy, where the same LLM is used for both patch generation and evaluation.
This design choice reduces computational overhead during the search process, and our experimental results indicate that self-evaluation contributes positively to the overall effectiveness of the repair strategy.

\textbf{Test-as-Judge}.
This strategy is designed for bug-fixing datasets with sufficient test cases (e.g., ConDefects), where each bug is associated with more than ten test cases that cover a wide range of scenarios and boundary conditions.
In this case, also supported by prior works~\cite{zhang2024appt,DBLP:conf/icse/YeMM22,chen2021evaluatinglargelanguagemodels}, we believe that relying on test execution results provides a highly reliable basis for evaluating patch quality.
Specifically, as shown in Equation~\ref{reward_equation2}, the reward $R$ is computed as the proportion of passed test cases, representing the test pass rate of the candidate patch:

\begin{equation} \label{reward_equation2}
R(a) = \frac{|\text{T}_{passed}|}{|\text{T}_{total}|}
\end{equation}

\begin{equation} \label{Qvalue2}
\mathbb{E}[R] = R
\end{equation}

\subsubsection{Search-State Updating} \label{stage-module-4}
In addition to using $R$ to immediately assess patch quality after each generation, \toolname{} also maintains state-level quality estimates throughout the entire search process.
The value of a patch depends not only on its own quality $R$ but also on the quality of the descendants it enables.
After reward $R$ is calculated for the generated patches, we update the Q-value of their parent patches using the following Equation~\ref{equation_back}: 

\begin{equation} \label{equation_back}
Q^{\prime}(a) = \beta \ \frac{\sum_{j=1}^{n} (Q_j \cdot N_j)}{\sum_{j=1}^{n} N_j} + (1- \beta)  \ Q(a) .
\end{equation}

Where $\beta$ is a forgetting factor that ranges from 0 to 1, and $N$ represents the number of children.
When $\beta$ is closer to 1, it indicates that the new value of Q is less influenced by the old value.
In our work, we set $\beta$ to 0.8.

Conceptually, the purpose of this stage is to aggregate path-level evidence over time.
A search state that consistently leads to better descendants should become more likely to be selected in later rounds, while states that repeatedly lead to poor descendants should gradually lose priority.

In each iteration, \toolname{} goes through the above four stages to search for and evaluate new patches, and then initiates the next round of searching based on the patches found and the evaluation results.
Upon completing all search iterations, we perform manual validation on the recorded plausible patches.
If they match the developer patches or are syntactically equivalent, we consider them as correct patches; otherwise, they are deemed wrong patches.

We present the pseudocode of \toolname{} as shown in the Algorithm~\ref{alg: mcts_algorithm}. We have annotated the range of each stage in the algorithm, and the comments describe what is done in each stage. 

\input{tab/mcts_algorithm}\label{algorithm}

\section{Experimental Setup}
\subsection{Research Questions}
We evaluate \toolname{} on the following research questions:
\begin{itemize}[leftmargin=*]
    \item \textbf{RQ1:} How does \toolname{} compare against the state-of-the-art APR techniques?
    \item \textbf{RQ2:} How does \toolname{} compare with using vanilla LLMs for APR? 
    \item \textbf{RQ3:} How much impact does each component of \toolname{} have on the overall effectiveness?
    \item \textbf{RQ4:} How effective is \toolname{} in fixing bugs across multiple languages and types?
    \item \textbf{RQ5:} How does \toolname{} perform on the vulnerability repair task?
    \item \textbf{RQ6:} How does the cost of \toolname{} compare to existing methods?
\end{itemize}

\subsection{Datasets}

We evaluate \toolname{} on five widely used benchmarks: QuixBugs~\cite{DBLP:conf/oopsla/LinKCS17}, Defects4J~\cite{DBLP:conf/issta/JustJE14}, ConDefects~\cite{wu2023condefectsnewdatasetaddress}, SWE-Bench-Lite~\cite{swe-bench}, and VUL4J~\cite{vul4j}. These datasets are commonly used in APR research~\cite{zhang2024survey,jiang2023impact,cao2025study,sweAgent} and cover multiple languages, defect types, and repair settings. We use the Java subset of QuixBugs (40 bugs), all 835 bugs in Defects4J, the Python subset of ConDefects, 300 issues in SWE-Bench-Lite, and 79 vulnerabilities in VUL4J.

\subsection{Baselines}
On QuixBugs and Defects4J, we compare \toolname{} against 12 state-of-the-art APR baselines from different categories, including five learning-based methods (i.e., SelfAPR~\cite{DBLP:conf/kbse/YeML0M22}, ITER~\cite{ye2024iter}, CURE~\cite{DBLP:conf/icse/JiangL021}, RewardRepair~\cite{DBLP:conf/icse/YeMM22}, and Recoder~\cite{DBLP:conf/sigsoft/ZhuSXZY0Z21}), two template-based methods (i.e., Repatt~\cite{jiang2023enhancingredundancybasedautomatedprogram} and GAMMA~\cite{zhang2023gammarevisitingtemplatebasedautomated}), and five LLM-based methods (i.e., RAPGen~\cite{DBLP:journals/corr/abs-2306-17077}, ChatRepair~\cite{DBLP:conf/issta/Xia024}, RepairAgent~\cite{bouzenia2024repairagentautonomousllmbasedagent}, GiantRepair~\cite{xu2025aligning}, D4C~\cite{li2025hybrid}). Additionally, we select 13 LLMs with varying parameter sizes as baselines, including five 3B models, six 7-9B models, and two API-accessible models.
On ConDefects, we compare \toolname{} against ChatRepair~\cite{DBLP:conf/issta/Xia024}, GPT-3.5, and AlphaRepair~\cite{xia2022less}. On SWE-Bench, we compare \toolname{} against Refact.ai Agent, SWE-agent~\cite{yang2024swe}, KGCompass~\cite{yang2025enhancing}, ChatRepair~\cite{DBLP:conf/issta/Xia024}, and OpenHands~\cite{wang2025openhands}. On VUL4J, we compare \toolname{} against FSV~\cite{effective}, NTR~\cite{ntr}, VRPILOT~\cite{vrpilot}, ChatRepair~\cite{DBLP:conf/issta/Xia024}, and APR4Vul~\cite{bui2024apr4vul}. For ChatRepair, we follow the original paper's configuration to reproduce its results on SWE-Bench-Lite and VUL4J.

\subsection{Evaluation Metrics}
We consider three widely used metrics~\cite{zhao2024enhancingautomatedprogramrepair,xin2024practicalusefulautomatedprogram,yang2024revisitingunnaturalnessautomatedprogram} to evaluate the effectiveness of both \toolname{} and baselines, and the quality of the generated patches.
The definitions of the metrics are listed as follows.

\begin{itemize}[leftmargin=*]
    \item Correct Fix (CF) is defined as the number of correctly fixed bugs, which can pass all the tests and is manually checked to ensure semantic or syntactic equivalence to the developer patch.
    \item Plausible Fix (PF) is defined as the number of bugs which can pass all the tests after fixing, while no further check is applied. 
    \item Repair Success Rate (RCR) represents the proportion of correctly fixed bugs among all bugs.
\end{itemize} 

\subsection{Implementation Details}

To implement \toolname{}, we use the API provided by OpenAI and the models available on Hugging Face for initialization.
We use tiktoken to count the number of tokens consumed in API calls and calculate the costs.
The temperature is set to 0.9, max\_token is set to 8000, and the patch size is set to 16.
For the primary model (GPT-3.5), we conduct extra experiments with the patch size set to 32.
The exploration constant is set to 0.7, alpha is set to 0.8, branch and max\_expansion is set to 1 and 3, respectively.
We implement \toolname{} based on the PyTorch and Transformers frameworks.
All experiments are conducted with two NVIDIA Tesla V100 GPUs on one Ubuntu 20.04 server.

\input{tab/baseline}


\section{Evaluation and Results}
\subsection{RQ1: Comparison With State-of-the-Arts}

\textbf{Experimental Design.} 
In RQ1, we aim to evaluate the performance of \toolname{}.
We consider 12 prior APR approaches and 13 LLMs as baselines.
To eliminate potential interference caused by model size, we select 7 best-performing models of different size and types to serve as the underlying model for \toolname{} in the subsequent experiments.

\textbf{Overall Performance}.
Table~\ref{tab:baseline-result} presents the comparison results of \toolname{} and baselines on Defects4J and QuixBugs benchmarks.
On the Defects4J dataset, \toolname{} obtains the highest 201 bug-fixes, fixing 21 more bugs than the second-place D4C.
Particularly, \toolname{} fixes 106 and 95 bugs on Defects4J-v1.2 and Defects4J-v2.
Although \toolname{} fixes 8 fewer bugs than ChatRepair on Defects4J-v1.2, it is acceptable given the \delete{performance} differences\revise{of patch size setting}.
ChatRepair generates and tests an average of 500 candidate patches per bug, while \toolname{} generates only 32 candidate patches per bug.
In addition, \toolname{} is able to provide more plausible fixes than previous studies.
Specifically, \toolname{} obtains a total of 280 plausible fixes, 70 more plausible fixes than that of D4C.
We list the number of project-level bug-fixes in Table~\ref{tab:best_result}.
When comparing \toolname{} against baseline methods, we find that the bug-fix distribution among the five methods shows considerable consistency.
\toolname{} significantly outperforms the other four baselines on Compress, JacksonDataBind, and Jsoup.
We also evaluate \toolname{} on the QuixBugs dataset.
The results show that \toolname{} is capable of fixing all the bugs in QuixBugs.

\input{tab/best_result_detail}

\begin{figure}[t]
\centering
    \subfigure[Venn on Defects4J-v1.2]{
        \includegraphics[width=0.45\columnwidth]{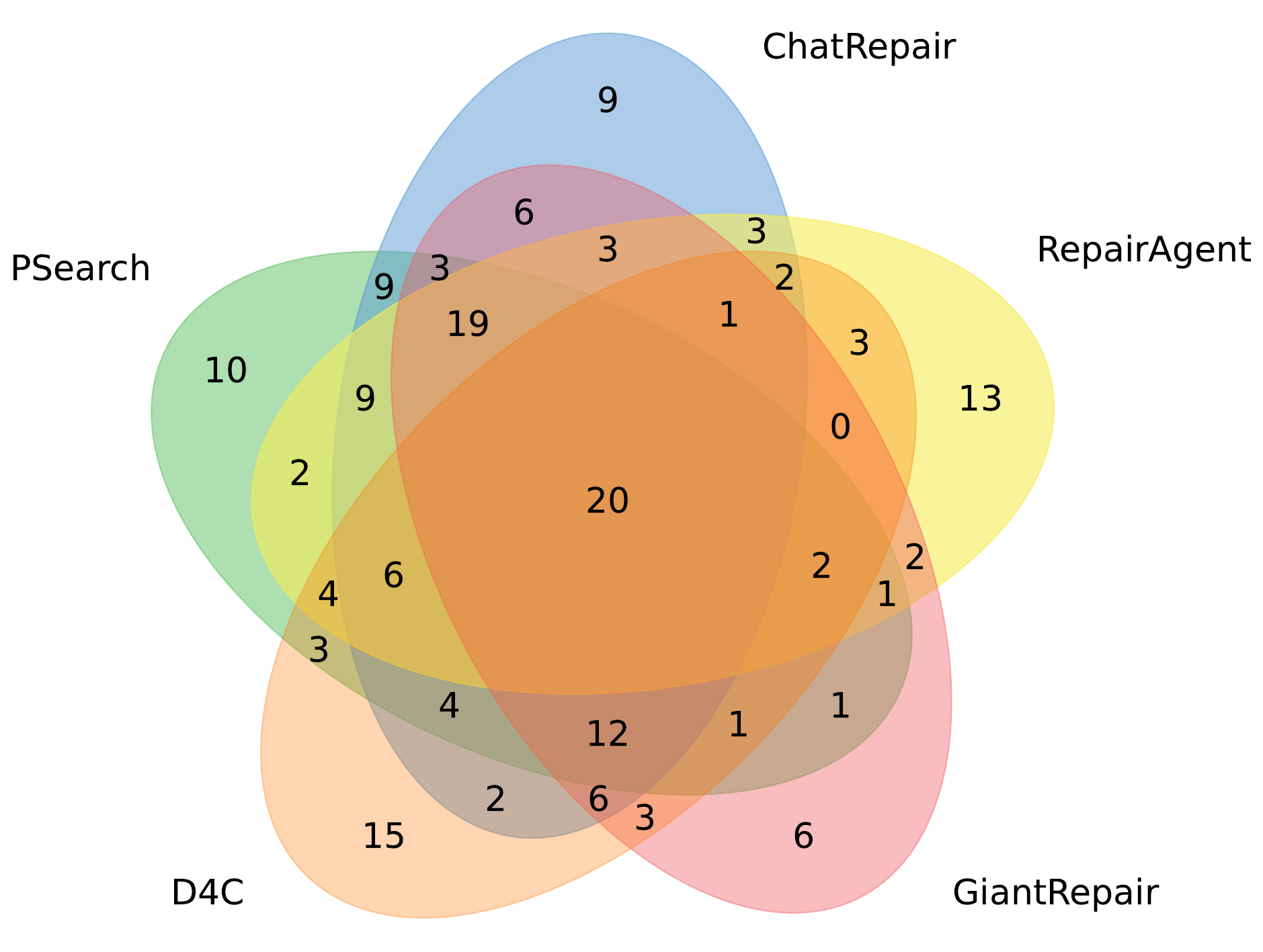}
        \label{best_venn_d4j12}
    }
    \subfigure[Venn on Defects4J-v2] {
        \includegraphics[width=0.45\columnwidth]{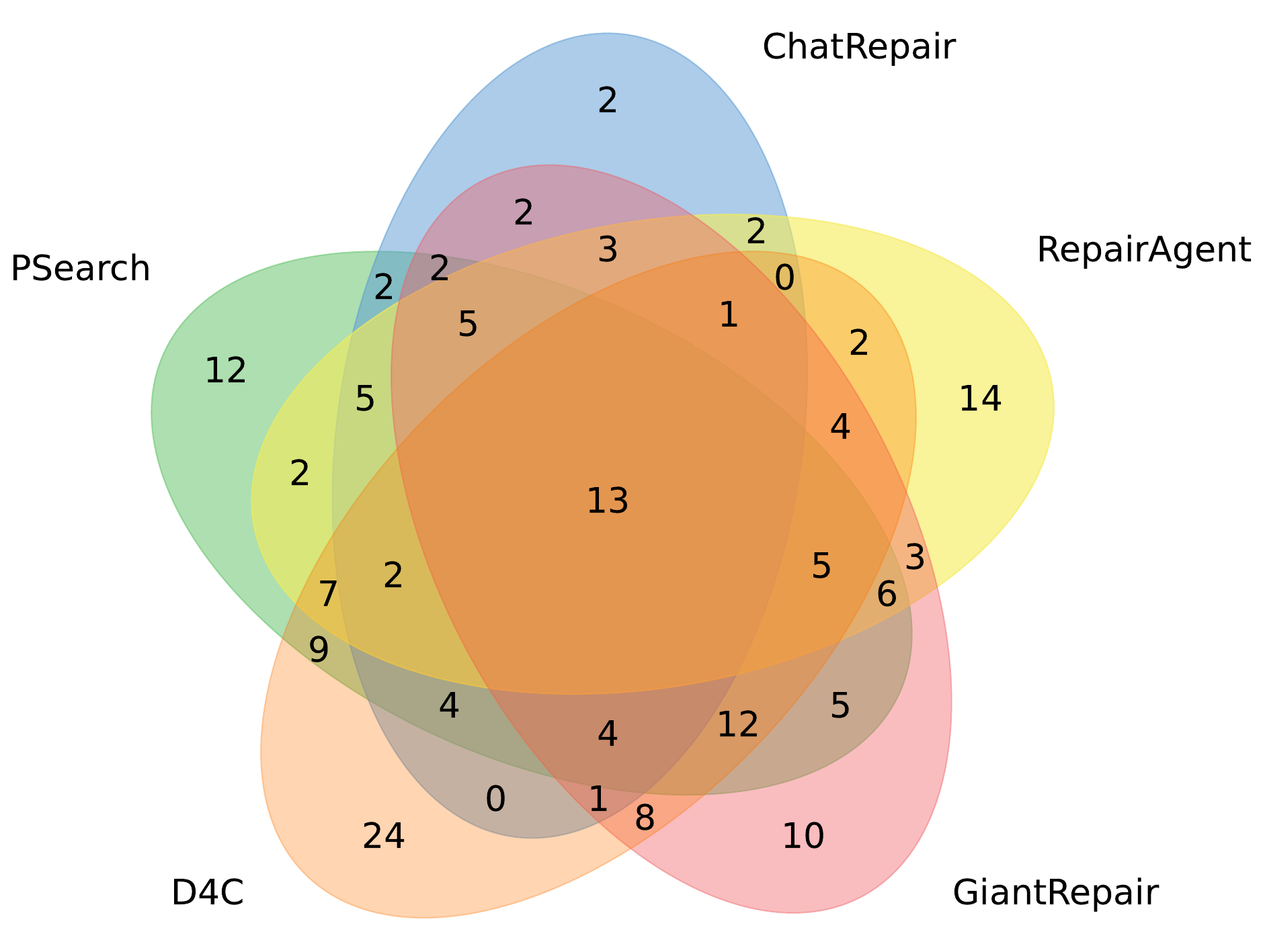}
        \label{best_venn_d4j2}
    }
    \caption{Bugfix Venn Diagram on Defects4J (\toolname{}, RepairAgent, ChatRepair, GiantRepair, D4C)}
    \label{best_d4j_venn}
\end{figure}

\textbf{Overlap Analysis}.
Among all 12 baselines, we select four state-of-the-art baselines for overlap analysis.
Figure~\ref{best_d4j_venn} presents the Venn diagram of the bugs fixed by D4C, GiantRepair, ChatRepair, RepairAgent, and \toolname{} on Defects4J-v1.2 and Defects4J-v2. It shows that \toolname{} maintains strong repair performance against these baselines. In particular, \toolname{} uniquely fixes 10 bugs on Defects4J-v1.2 and 12 bugs on Defects4J-v2 that are not repaired by any of the other four methods, demonstrating that \toolname{} can substantially complement existing repair approaches. Meanwhile, the five methods still share a non-trivial overlap, with 20 commonly fixed bugs on Defects4J-v1.2 and 13 on Defects4J-v2, indicating that they capture some common repair patterns while retaining different strengths.

\begin{figure}[t!]
\centering
\includegraphics[width=\columnwidth]{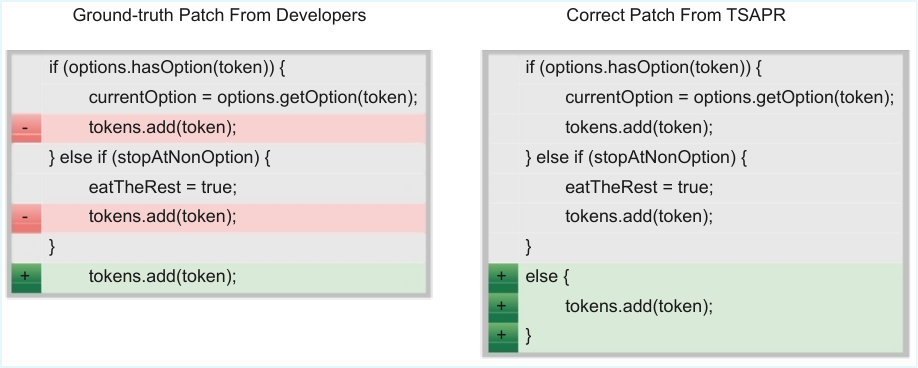}
\caption{Unique Fix (Cli\_19) from \toolname{}}  \label{unique_fix}
\end{figure} 

\textbf{Case Study}.
To better illustrate the advancement of \toolname{}, we provide several notable fixes.
\toolname{} can fix both Gson\_15 and Lang\_16 which ChatRepair~\cite{DBLP:conf/issta/Xia024} mentions as unique fixes.
We further demonstrate a unique fix from \toolname{} results, as shown in Figure~\ref{unique_fix}.
Cli\_19 is a function-level bug from Defects4J-v2, which cannot be fixed by simply replacing one or several buggy lines.
Instead, fixing this bug requires modifying the function in multiple places, thus bringing much difficulty to APR and no baselines can fix it.
The key to fixing Cli\_19 lies in understanding that the action $tokens.add(token)$ is necessary under all conditional branches.
As shown in Figure~\ref{unique_fix}, \toolname{} arrives at a correct patch that is different from the developer patch but semantically equivalent. 

\finding{1}{
\toolname{} significantly outperforms all prior APR methods on plausible/correct fixes, with 106 bug-fixes on Defects4J-v1.2, 95 bug-fixes on Defects4J-v2 and 40 bug-fixes on QuixBugs.
}

\subsection{RQ2: Comparison with LLMs}

\textbf{Experimental Design.}
In RQ1, we have demonstrated that \toolname{} achieves impressive performance across a range of APR techniques and LLMs.
In RQ2, we further investigate the extent to which \toolname{} improves performance across different underlying LLMs, and whether these improvements are attributable to our proposed framework rather than to the inherent capabilities of the models themselves.
To this end, we select seven of the best-performing LLMs from each model scale category in RQ1 and apply our framework to them.

\input{tab/Defects4jResult}
\textbf{Results and Analysis.}
Table~\ref{vanilla} presents the performance improvements achieved by \toolname{} across different underlying models.
Results show that the repair capabilities of all seven LLMs generally improve after applying \toolname{}.
Among these, Yi-Coder-9B, Qwen2.5-Coder-7B, GPT-4o-mini and GPT-3.5 demonstrate the most significant improvements, with an increase of 37, 28, 30 and 27 bug-fixes, respectively.
Moreover, with the patch size set to 32, GPT-3.5 (\toolname{}) can fix 201 bugs, which is 69 more bug-fixes than vanilla GPT-3.5.
Llama-3.1-8B and Qwen2.5-Coder-3B show certain improvement, both with an additional 9 bug-fixes. 

In terms of buggy types, the success rate for fixing single-line (SL) and single-hunk (SH) bugs is significantly higher than that for single-function (SF) bugs.
For the former two types of bugs, LLMs can pinpoint the exact location of buggy lines, and the logic of the buggy programs is relatively simpler, requiring less modification compared to SF bugs.
Thus it is harder for LLMs to fix SF bugs.
Compared to Vanilla LLMs, we notice that \toolname{} significantly enhances the effectiveness of LLMs in fixing SF bugs, with GPT-4o-mini fixing 12 more SF bugs, GPT-3.5 fixing 8 more SF bugs, Yi-Coder-9B fixing 14 more SF bugs, Qwen2.5-Coder-7B fixing 8 more SF bugs, and Qwen2.5-Coder-3B fixing 4 more SF bugs.
It indicates that \toolname{} has a particular advantage in fixing complex bugs.

\finding{2}{
The comparison results between \toolname{} and vanilla LLMs show that, with the same patch size (e.g., 16) and backbone model, \toolname{} can improve the repair effectiveness on Defects4J by over 20\% compared to vanilla LLMs, e.g., improving GPT-3.5 by 20.45\% (132 $\rightarrow$ 159), improving GPT-4o-mini by 23.43\% (128 $\rightarrow$ 158).
}

\subsection{RQ3: Effectiveness of Each Component}
\textbf{Experimental Design.} 
In RQ3, we perform ablation study to validate the effectiveness of each component, including test information, prompting strategy and search/evaluation.
We incrementally incorporate each component into our method to see its impact on performance.

\subsubsection{RQ3.1: Effectiveness of Test Information}
\input{tab/without_test}
Table~\ref{without_test} compares the repair performance with and without test information. The results show that test information consistently improves the effectiveness of all seven evaluated LLMs. In particular, GPT-4o-mini and GPT-3.5 benefit the most, fixing 21 and 18 more bugs, respectively, when test information is provided. Yi-Coder-9B and Qwen2.5-Coder-3B improve by 12 bugs, while Stable-Code-3B and Llama-3.1-8B improve by 9 bugs, and Qwen2.5-Coder-7B improves by 8 bugs. These results indicate that test information provides useful repair signals for patch generation, helping the models better understand bug behavior and produce more accurate fixes. Overall, test information is an effective component for improving LLM-based APR.

\subsubsection{RQ3.2: Comparison of Prompting Strategies}
We compare three prompt strategies for patch generation: Vanilla prompting, Chain-of-Thought (CoT), and Tree-of-Thought (ToT). As shown in Table~\ref{cot_tot}, CoT is generally the most effective strategy. Compared with Vanilla prompting, most LLMs achieve better repair performance with CoT. Yi-Coder-9B and GPT-3.5 show the largest gains, with CF improving by 31 and 7, and PF improving by 32 and 10, respectively. In contrast, ToT does not consistently improve performance and even reduces CF for GPT-4o-mini, Llama-3.1-8B, and Stable-Code-3B by 7, 19, and 2, respectively. Overall, CoT outperforms ToT on most evaluated LLMs.
\input{tab/cot_tot}

\subsubsection{RQ3.3: Effectiveness of Search and Evaluation}

\input{tab/mcts_result}

To evaluate the impact of the search and evaluation components in \toolname{}, we compare it with CoT-enhanced and vanilla LLM baselines. As shown in Table~\ref{detailed_results}, \toolname{} consistently improves the effectiveness of all seven evaluated LLMs over both CoT and Vanilla prompting. In particular, GPT-3.5, GPT-4o-mini, and Yi-Coder-9B with \toolname{} fix 20, 27, and 6 more bugs than their CoT counterparts, respectively. When the patch size increases to 32, GPT-3.5 with \toolname{} fixes 42 additional bugs.
We further observe that larger models, such as GPT-4o-mini, GPT-3.5, Yi-Coder-9B, and Qwen2.5-Coder-7B, benefit more from \toolname{} than smaller ones. For GPT-4o-mini and GPT-3.5, 90\% (27/30) and 74\% (20/27) of the overall improvement over Vanilla LLMs can be attributed to search and evaluation, respectively. This suggests that stronger models can provide more accurate patch evaluation, which in turn guides the search more effectively.
In addition, the advantage of \toolname{} becomes more evident as patch size increases. For Llama-3.1-8B and Qwen2.5-Coder-3B, \toolname{} is initially comparable to or slightly worse than CoT at small patch sizes, but gradually matches and then surpasses CoT as the patch size grows. This indicates that search and evaluation are especially helpful for handling more complex bugs.

\textbf{Effectiveness of Large Patch Size.}
To further investigate the impact of large patch size, we select GPT-3.5 for extreme testing.
We increase the patch size from 32 to 500 (50 iterations, 10 patches per iteration) to align with ChatRepair's configuration.
We list the newly fixed bugs in Table~\ref{extra_fix_patch_500}, where $\checkmark$ represents a correct fix, and \texttimes \ represents a plausible but not correct fix.
It can be observed that a larger patch size (500) leads to more plausible fixes (16) and correct fixes (11).
However, as the patch size increases, the number of newly fixed bugs significantly decreases.
This indicates that \toolname{} has already approached its upper limit.

\input{tab/patchsize_500_extra_fix}

\finding{3}{
All components, including test information, CoT prompting, search, and evaluation, have a positive effect on \toolname{}. 
Among them, test information is effective for all LLMs (e.g., helping GPT-4o-mini fix 21 more bugs). CoT is effective for 6/7 LLMs (e.g., helping Yi-Coder-9B fix 31 more bugs). Search and evaluation are effective for all LLMs (e.g., helping GPT-4o-mini repair 30 more bugs). 
Moreover, large-scale models provide more accurate evaluations of patch quality, leading to better search results (e.g., GPT-3.5 fixes 27 more bugs, while Qwen2.5-Coder-3B only fixes 8 more bugs). Extreme testing shows a larger patch size (32 $\rightarrow$ 500) helps \toolname{} fix 11 more bugs, suggesting that increased search budget further enhances its repair effectiveness.
}

\subsection{RQ4: Performance on Multi-lingual and Multi-type Bugs}

\textbf{Experimental Design.}
In RQ 1-3, we have validated the effectiveness of \toolname{} on project-level Java bugs (e.g., Defects4J).
To further validate the repair capability of \toolname{} on bugs of different types and in different languages, we perform extra experiments on the ConDefects-Python and the SWE-Bench dataset.
For ConDefects, we compare \toolname{} with ChatRepair, GPT-3.5 and AlphaRepair.
To ensure fairness, we follow ChatRepair and employ GPT-3.5 as the experimental LLM.
For SWE-Bench, we choose the open-source model Qwen3-Coder-480B as the base model, as it has been demonstrated by prior work to perform well on SWE-Bench and offers a low cost.

\textbf{Results and Analysis.}
We first report the results of \toolname{} on ConDefects.
As shown in Table~\ref{tab:condefects_result}, when patch size = 48 (16 iterations, 3 patches per iteration), \toolname{} obtains 211 plausible fixes and 204 correct fixes, which is 40 more plausible fixes and 39 more correct fixes than ChatRepair.
Since the patch size for ChatRepair is set to 500, it can be seen that with less than one-tenth of the patch size, \toolname{} still significantly enhances the patch search performance of LLMs.
When we increase search iteration to 32 and set patch size to 96, we find that the performance of \toolname{} is further enhanced, with 287 plausible fixes and 264 correct fixes, which surpasses ChatRepair by 23/38 correct/plausible fixes.
Additionally, we find that Test-as-Judge enables LLMs to quickly generate patches that satisfy simple test cases, and then iteratively refine the details of the patches through complex test cases until all boundary conditions are met.
Compared to allowing the model to search patches without evaluation, Test-as-Judge guides LLMs in the right direction for repairs, improving the efficiency of patch search.

\input{tab/Condefects}

On SWE-Bench-Lite, we use the open-source Qwen3-Coder-480B as the base model.
We use the same configuration as Defects4J to set the patch size to 16 and score patches by test reports and patch content.
We directly use the test cases and perfect localization provided in the dataset for the convenience of evaluating the patch search capability of \toolname{}.
As shown in Table~\ref{tab:swe_result}, compared with vanilla LLMs, \toolname{} helps Qwen3-Coder-480B fix 51 more bugs.
Compared to ChatRepair, \toolname{} fixes 35 more bugs. It suggests that effective patch search is critical for scaling APR to realistic software engineering tasks.
By evaluating intermediate search states and allocating search effort to more promising trajectories, \toolname{} can substantially improve the repair capability of the underlying base model.
In addition, \toolname{} outperforms recent approaches such as KGCompass~\cite{yang2025enhancing} and OpenHands~\cite{wang2025openhands}.
In future work, \toolname{} can be integrated with advanced fault localization and test generation tools~\cite{zhang2025improving,zhang2025improvingtosem,zhang2024testbench} to form agent-based frameworks with powerful repair capabilities.
\input{tab/swe_result}

The above experimental results demonstrate that \toolname{} has significant advantages over previous methods and vanilla LLMs in repairing bugs across multiple languages (Java/Python) and multiple types (Repository/Competition).

\finding{4}{
\toolname{} demonstrates excellent performance in multi-language and multi-type bug repair, successfully fixing 164 out of 300 issues on the repository-level defect dataset SWE-Bench-Lite, ranking third among all five baselines. Moreover, \toolname{} fixes 264 bugs in the competition-level Python defect dataset ConDefects, which is 23 more than the second-best ChatRepair.
}

\subsection{RQ5: Performance of \toolname{} on Vulnerability Repair}
\textbf{Experimental Design.}
In the previous RQs, we have validated \toolname{}'s repair performance across various types of defects.
In RQ5, we further test \toolname{}'s capability in vulnerability repair to evaluate its generality.
Thus, we extend our framework to the vulnerability repair domain and develop \toolname{}-Vul.
We use the same configuration as in the previous experiments and adopt GPT-3.5 as the base model.

\textbf{Results and Analysis.}
As shown in Table~\ref{tab:vul}, we report the number of vulnerabilities fixed by \toolname{} and baseline methods on VUL4J.
$FSV_{Codex}$ denotes FSV with zero-shot Codex model and $FSV_{finetuned}$ denotes FSV fine-tuned with general APR data.
\toolname{} successfully fixes a total of 27 vulnerabilities, including 4 multi-method vulnerabilities, surpassing all baseline methods, achieving a repair success rate of 34.17\%.
When using the same base model (GPT-3.5), \toolname{} fixes 12 more vulnerabilities than ChatRepair, demonstrating its superiority in vulnerability repair.
Notably, APR4Vul has shown that ten mainstream traditional repair tools collectively fix only 16 vulnerabilities on VUL4J, which clearly demonstrates the limitations of conventional repair methods in addressing vulnerabilities.
In contrast, \toolname{} breaks through this barrier, demonstrating strong generalization capability.

\input{tab/vul4j}

\finding{5}{
\toolname{} successfully fixes 27 out of the 79 vulnerabilities in VUL4J, outperforming all baselines and demonstrating strong generalization capability.
}

\subsection{RQ6: Cost Analysis}
\textbf{Experimental Design.} 
In RQ6, we aim to analyze the differences between \toolname{} and existing APR tools in terms of patch size, time, token consumption, and monetary cost.
Specifically, we select ChatRepair and RepairAgent as baselines, and use the cost on Defects4J for comparison.

\input{tab/cost}

\textbf{Results and Analysis.}
The comparison result is shown in Table~\ref{tab:cost_analysis}.
With the patch size set to 32, which is the smallest among all three baselines, \toolname{} spends an average of 50 minutes per bug, shorter than that of ChatRepair.
Moreover, \toolname{} also has a significant advantage in terms of the average number of tokens spent and monetary cost per bug, which is only 19\% of the 210,000 tokens reported by ChatRepair and 14.8\% of the 270,000 tokens reported by RepairAgent.
In terms of pricing, we calculate based on the current API price.
The cost of \toolname{} is \$0.06 per bug, which is 43\% of ChatRepair (\$0.14) and RepairAgent (\$0.14).

\finding{6}{
\toolname{} proves low cost and high performance efficiency, taking an average of 50 minutes and \$0.06 per bug, which is only 16.7\% and 43\% of baselines.
}

\section{Threats to Validity}
\label{sec:threats}

\textbf{Internal Threat.}
The main internal threat involves potential data leakage, implementation bias, and stochasticity in LLM-based generation.
To address this, we assess data leakage from multiple perspectives, including analyzing training data of open-source models when available, incorporating additional benchmarks with different origins, and examining overlap between generated patches and developer patches to detect potential memorization effects.
Although we do not observe strong evidence of leakage, we cannot fully rule out subtle contamination due to the scale of modern pretraining corpora.
In addition, randomness in LLM decoding and the search procedure (e.g., MCTS exploration) may introduce variance across runs.
To mitigate this, we aggregate results using repeated evaluations and expected reward estimation, which improves stability.

\textbf{External Threat.}
The main external threat to validity is whether the performance of \toolname{} generalizes beyond the evaluated datasets and settings.
To mitigate this concern, we evaluate \toolname{} on both repository-level benchmarks (Defects4J, SWE-Bench-Lite) and competition-level or synthetic benchmarks (ConDefects, VUL4J), covering multiple programming languages and defect types.
This helps ensure that improvements are not limited to a specific dataset style or bug distribution.
Moreover, \toolname{} is designed to be agnostic to programming languages and bug categories, suggesting applicability beyond Java and Python.
However, real-world software systems may present additional challenges that are not fully captured by existing benchmarks.
Therefore, while our results are consistent across benchmarks, absolute performance in industrial settings may differ.

\section{Conclusion}
In this paper, we propose \toolname{}, a search-based patch generation framework for LLM-based APR. \toolname{} improves patch generation by maintaining structured search states, evaluating intermediate search paths, and iteratively refining promising candidates. Rather than relying on unguided trial-and-error generation, it formulates patch generation in APR as an evaluate-and-improve process. We implement \toolname{} with MCTS, APR-specific reasoning, self-reflection, and adaptive patch evaluation.
Experiments on 835 bugs from Defects4J show that \toolname{} fixes 201 bugs and outperforms 12 state-of-the-art baselines. We further validate its multilingual and multi-type repair ability on ConDefects-Python and SWE-Bench-Lite, and its scalability on VUL4J. Compared with existing LLM-based APR tools, \toolname{} is faster and reduces monetary cost by more than 50\%. These results indicate that progress in LLM-based APR depends not only on stronger generation models, but also on more effective search-based patch generation that can prioritize promising search trajectories early.

\section*{Data Availability Statement}
The artifact, including code, scripts, data and results, is available at \textit{https://github.com/Tomsawyerhu/Psearch}.

\section*{Acknowledgments}
This research was supported in part by Natural Science Foundation of Jiangsu Province (BK20251458), and Fundamental Research Funds for the Central Universities (AE89991/463).

\bibliographystyle{ACM-Reference-Format}
\bibliography{reference}

\end{document}

%% file: tab/mcts_algorithm.tex
\begin{algorithm}[t!]
\small
\caption{The proposed \toolname{} algorithm }
\renewcommand{\algorithmicrequire}{\textbf{Input:}}
\renewcommand{\algorithmicensure}{\textbf{Output:}}

\label{alg: mcts_algorithm}
\begin{algorithmic}[1]
\REQUIRE {
orig$\_$buggy$\_$program $bp_o$, 
orig$\_$test$\_$result $t_o$, 
patch$\_$generator $\pi_p$, 
patch$\_$evaluator $\pi_r$, 
max$\_$round $T$,
branch $b$,
exploration$\_$constant $\epsilon$,
quit$\_$at$\_$first$\_$plausible $quit$.}
\STATE $T_q \leftarrow $Initialize$\_$Tree$(bp_o,t_o)$
\STATE $\pi_p, \pi_r \leftarrow $Initialize$\_$Models$(\pi_\theta, \pi_\theta)$
\FOR {$i$ in range$(T)$}
\STATE $C\leftarrow $root$(T_q)$
\STATE \textit{\textbf{---------------Patch\ Selection---------------}}
\WHILE{$C$ is not leaf node}
\STATE $C\leftarrow argmax_{C^{'}\in \text{children}(C)}(\bar{X}_C+\epsilon\sqrt{\frac{\ln{n_C}}{n_{C^{'}}}})$  
\ENDWHILE

\STATE \textit{\textbf{---------------Patch\ Generation---------------}}
\STATE $bp_{old},t_{old} \leftarrow$Extract$\_$Bug$\_$Info($C$) 
\FOR{$j$ in range($b$)}
\STATE $cot\leftarrow$Get$\_$CoT($\pi,bp,t$) 
\STATE $P_j,ref\leftarrow$Repair($\pi,bp_{old},t_{old},cot$)
\STATE $t_{new}\leftarrow$Validate$\_$Patch($P_j$)
\IF{Is$\_$Pass($t_{new}$) \AND $quit$}
\STATE \textbf{Return} $P_j$
\ELSE
\STATE \textit{\textbf{---------------Patch Evaluation---------------}}
\STATE $r\leftarrow$Get$\_$Reward($\pi_r,P_j,t_{new},cot,ref$)
\STATE $b_{new}\leftarrow$Construct$\_$Bug($bp_{old},P_j$)
\STATE $T_q\leftarrow$Update$\_$Tree($b_{new},t_{new}$)
\ENDIF
\ENDFOR

\STATE \textit{\textbf{---------------Search State Updating---------------}}
\STATE Back$\_$Propagate($C,r$)
\ENDFOR
\STATE $P$=Get$\_$Plausible$\_$Patches($T_q$)
\STATE \textbf{Return} $P$
\ENSURE $P$
\end{algorithmic}
\end{algorithm}

%% file: tab/baseline.tex
\begin{table*}[htbp]
\footnotesize
    \centering
    \caption{Comparison with baselines on Defects4J and QuixBugs (correct/plausible fix).}
    \label{tab:baseline-result}
    \resizebox{\linewidth}{!}{
    \begin{tabular}{c|l|cc|ccc|c} 
        \toprule
         & Method & Model & Patch Size & Defects4J-v1.2 & Defects4J-v2 & Total & QuixBugs \\
        \midrule
        
        \multirow{9}{*}{APR} & SelfAPR~\cite{DBLP:conf/kbse/YeML0M22}    & T5                 & 150 & 65/74           & 45/47           & 110/121 & -             \\ 
        & \revise{ITER}~\cite{ye2024iter}      & \revise{T5}                   & \revise{1000} & \revise{59/89}             & \revise{19/36}           & \revise{78/125}  & \revise{-}            \\ 
        & CURE~\cite{DBLP:conf/icse/JiangL021}      & GPT-2                   & 5000 & 57/-             & 19/-           & 76/-  & 26            \\ 
        & RAPGen~\cite{DBLP:journals/corr/abs-2306-17077}      & CodeT5                  & $\leq$100 & 72/-             & 53/-           & 125/- & -             \\ 
        & RewardRepair~\cite{DBLP:conf/icse/YeMM22}         & Transformer       & 200 & 45/-             & 45/-           & 90/-  & 20            \\ 
        & Recoder~\cite{DBLP:conf/sigsoft/ZhuSXZY0Z21}         & TreeGen              & 100 & 53/-             & 19/-           & 72/-  & 31            \\ 
        & Repatt ~\cite{jiang2023enhancingredundancybasedautomatedprogram}  & NA & 1200 & 40/70& 35/68 & 75/138 & - \\
        & GAMMA~\cite{zhang2023gammarevisitingtemplatebasedautomated}  & GPT-3.5 & 250 & 82/108 & 45/- & 127/- & 22 \\
        & ChatRepair~\cite{DBLP:conf/issta/Xia024} & GPT-3.5 & $\leq$500 & 114/- & 48/- & 162/- & 40 \\
        & RepairAgent ~\cite{bouzenia2024repairagentautonomousllmbasedagent} & GPT-3.5 & 117 & 74/96 & 90/90 & 164/186 & - \\
        & GiantRepair~\cite{li2025hybrid} & GPT-3.5 & $\leq$200 & 86/- & 84/- & 170/- & - \\
        & D4C~\cite{xu2025aligning} & GPT-4 & 10 & 84/96 & 96/114 & 180/210 & - \\
        \midrule
        \multirow{13}{*}{LLM} & Stable-Code-3B & Stable-Code-3B & 16 & 31/49 & 27/50 & 58/99 & 20 \\
        & Calme-3.1-3B & Calme-3.1-3B & 16 & 25/44 & 20/42 & 45/86 & 19 \\
        & Starcoder2-3B & Starcoder2-3B & 16 & 19/35 & 24/44 & 43/79 & 18 \\
        & Qwen2.5-Coder-3B & Qwen2.5-Coder-3B & 16 & 44/68 & 43/70 & 87/138 & 27 \\
        & Llama-3.2-3B & Llama-3.2-3B & 16 & 32/53 & 27/42 & 59/95 & 21 \\  
        & Phi-3.5-mini & Phi-3.5-mini & 16 & 28/52 & 29/53 & 57/105 & 19 \\
        & DeciLM-7B & DeciLM-7B & 16 & 23/42 & 22/41 & 45/83 & 19 \\
        & Falcon-7B & Falcon-7B & 16 & 8/21 & 10/25 & 18/46 & 4 \\
        & Yi-Coder-9B & Yi-Coder-9B & 16 & 48/73 & 58/93 & 106/166 & 31 \\  
        & Llama-3.1-8B & Llama-3.1-8B & 16 & 43/71 & 43/68 & 86/139 & 25 \\  
        & Qwen2.5-Coder-7B & Qwen2.5-Coder-7B & 16 & 38/66 & 41/70 & 79/132 & 25 \\
        & GPT-4o-mini & GPT-4o-mini & 16 & 67/89 & 61/81 & 128/170 & 35 \\  
        & GPT-3.5 & GPT-3.5 & 16 & 69/92 & 63/84 & 132/176 & 36 \\
        \midrule
        
        \multirow{2}{*}{Ours} &  \cellcolor{gray!10} \toolname{}    & \cellcolor{gray!10}GPT-3.5     & \cellcolor{gray!10}16 & \cellcolor{gray!10}86/112              &   \cellcolor{gray!10}73/104         & \cellcolor{gray!10}159/216  &  \cellcolor{gray!10}40             \\ 
        & \cellcolor{gray!20} \toolname{}   & \cellcolor{gray!20}GPT-3.5     & \cellcolor{gray!20}32 & \cellcolor{gray!20}106/146              &   \cellcolor{gray!20}95/134         & \cellcolor{gray!20}201/280  &  \cellcolor{gray!20}40             \\ 
        \bottomrule
    \end{tabular}}
\end{table*}

%% file: tab/best_result_detail.tex
\begin{table*}[htbp]
  \centering
\caption{Results of \toolname{} (GPT-3.5, 32 patch) on Defects4J. Core is short for JacksonCore, Xml is short for JacksonXml, Databind is short for JacksonDatabind, Collect is short for Collections.}
\label{tab:best_result}
\tabcolsep=2pt
    \resizebox{\linewidth}{!}{
    \begin{tabular}{c|ccccccccccccccccc|c}
    \toprule
     \toolname{} & Closure & Chart & Lang  & Math & Mockito & Time & Cli & Codec&  Collect & Compress & Csv & Gson & Core & Databind & Xml & JxPath & Jsoup & Total\\
    \midrule
    \# Bugs & 174 & 26 & 64 & 106 & 38 & 26 & 39 & 18 & 4 & 47 & 16 & 18 & 26 & 112 & 6 & 22 & 93 & 835 \\
    \midrule
    Plausible & 45 & 13 & 29 & 45 & 8 & 6 & 14 & 8 & 0 & 23 & 8 & 6 & 5 & 31 & 1 & 3 & 35 & 280 \\
    Correct & 28 & 12 & 24 & 32 & 8 & 4 & 12 & 5 & 0 & 15 & 7 & 4 & 4 & 18 & 1 & 1 & 26 & 201 \\
    \midrule
    RepairAgent & 27 & 11 & 17 & 29 & 6 & 2 & 8 & 9 & 1 & 10 & 6 & 3 & 5 & 11 & 1 & 0 & 18 & 164 \\
    ChatRepair & 37 & 15 & 21 & 32 & 6 & 3 & 5 & 8 & 0 & 2 & 3 & 3 & 3 & 9 & 1 & 0 & 14 & 162 \\
    GiantRepair & 34 & 8 & 13 & 26 & 6 & 1 & 7 & 8 & 0 & 12 & 6 & 6 & 8 & 15 & 1 & 1 & 18 & 170 \\
    D4C & 28 & 5 & 25 & 18 & 6 & 4 & 13 & 3 & 0 & 11 & 7 & 3 & 6 & 24 & 1 & 1 & 25 & 180 \\
    \bottomrule
    \end{tabular}
    }
\end{table*}

%% file: tab/Defects4jResult.tex
\begin{table*}[htbp]
\centering
\caption{Comparison of correct/plausible fix between Vanilla LLMs and \toolname{} on Defects4J and QuixBugs, including three types of bugs, single-line (SL), single-hunk (SH) and single-function (SF).} \label{vanilla}
\resizebox{0.9\linewidth}{!}{
\begin{tabular}{c|lc|cccc|c}
\toprule
Category & Model & Patch Size & SL & SH & SF & Defects4J & QuixBugs\\
\midrule
\multirow{4}{*}{3B} 
& Qwen2.5-Coder-3B& 16 & 56/79 & 13/25 & 18/34 & 87/138 & 27 \\ 
& \cellcolor{gray!20}{Qwen2.5-Coder-3B (\toolname{})} & \cellcolor{gray!20}{16} & \cellcolor{gray!20}{60/86} & \cellcolor{gray!20}{13/24} & \cellcolor{gray!20}{22/43} & \cellcolor{gray!20}{95/153} & \cellcolor{gray!20}{-} \\
& Stable-Code-3B & 16 & 39/56 & 4/12 & 15/31 & 58/99 & 20\\
& \cellcolor{gray!20}{Stable-Code-3B (\toolname{})} & \cellcolor{gray!20}{16} & \cellcolor{gray!20}{40/58} & \cellcolor{gray!20}{5/13} & \cellcolor{gray!20}{17/35} & \cellcolor{gray!20}{62/106} & \cellcolor{gray!20}{-}\\
\midrule
\multirow{6}{*}{7B-9B} 
& Yi-Coder-9B & 16 & 60/77 & 16/30 & 30/59 & 106/166 & 31\\
& \cellcolor{gray!20}{Yi-Coder-9B (\toolname{})} & \cellcolor{gray!20}{16} &\cellcolor{gray!20}{73/90} & \cellcolor{gray!20}{26/37} & \cellcolor{gray!20}{44/63} & \cellcolor{gray!20}{143/190} & \cellcolor{gray!20}{-}\\
& Llama-3.1-8B & 16 & 48/63 & 12/21 & 26/55 & 86/139 & 25\\
& \cellcolor{gray!20}{Llama-3.1-8B (\toolname{})} & \cellcolor{gray!20}{16} & \cellcolor{gray!20}{54/75} & \cellcolor{gray!20}{14/26} & \cellcolor{gray!20}{27/61} & \cellcolor{gray!20}{95/162} & \cellcolor{gray!20}{-}\\
& Qwen2.5-Coder-7B & 16 & 46/62 & 11/18 & 22/52 & 79/132 & 25\\
& \cellcolor{gray!20}{Qwen2.5-Coder-7B (\toolname{})} & \cellcolor{gray!20}{16} & \cellcolor{gray!20}{61/78} & \cellcolor{gray!20}{16/34} & \cellcolor{gray!20}{30/59} & \cellcolor{gray!20}{107/171} & \cellcolor{gray!20}{-}\\
\midrule
\multirow{5}{*}{API} 
& GPT-4o-mini & 16 & 65/72 & 27/37 & 36/61 & 128/170 & 35\\
& \cellcolor{gray!20}{GPT-4o-mini (\toolname{})} & \cellcolor{gray!20}{16} &\cellcolor{gray!20}{78/92} & \cellcolor{gray!20}{32/45} & \cellcolor{gray!20}{48/71} & \cellcolor{gray!20}{158/208} & \cellcolor{gray!20}{40}\\
& GPT-3.5 & 16 & 67/73 & 29/38 & 36/65 & 132/176 & 36\\
& \cellcolor{gray!20}{GPT-3.5 (\toolname{})} & \cellcolor{gray!20}{16} &\cellcolor{gray!20}{84/96} & \cellcolor{gray!20}{31/46} & \cellcolor{gray!20}{44/74} & \cellcolor{gray!20}{159/216} & \cellcolor{gray!20}{40}\\

& \cellcolor{gray!20}{GPT-3.5 (\toolname{})} & \cellcolor{gray!20}{32} & \cellcolor{gray!20}{104/121} & \cellcolor{gray!20}{42/64} & \cellcolor{gray!20}{55/95} & \cellcolor{gray!20}{201/280} & \cellcolor{gray!20}{40}\\

\bottomrule
\end{tabular}}
\end{table*}

%% file: tab/without_test.tex
\begin{table*}[htp]
    \centering
    \caption{Comparison of the number of bugs-fixes with test information vs. without test information.}\label{without_test}
    \begin{tabular}{lccccccc}
        \toprule
        &Qwen2.5-Coder-3B & Stable-Code-3B & Yi-Coder-9B & Llama-3.1-8B & Qwen2.5-Coder-7B  & GPT-4o-mini & GPT-3.5\\ 
        \midrule
        \textbf{w/o test} & 75 & 49 & 94 & 77 & 71 & 107   & 114 \\ 
         \textbf{w/ test}  & 87(\textcolor{red}{$\uparrow$ 12}) & 58(\textcolor{red}{$\uparrow$ 9}) & 106(\textcolor{red}{$\uparrow$ 12}) & 86(\textcolor{red}{$\uparrow$ 9}) & 79(\textcolor{red}{$\uparrow$ 8}) & 128(\textcolor{red}{$\uparrow$ 21}) & 132(\textcolor{red}{$\uparrow$ 18})  \\

        \bottomrule
    \end{tabular}
\end{table*}

%% file: tab/cot_tot.tex
\begin{table}[t]
    \centering
    \caption{CF under different prompting strategies.}
        \begin{tabular}{lccc}
            \toprule
            Model & Vanilla & CoT & ToT \\
            \midrule
            GPT-4o-mini & 128 & 131(\textcolor{red}{$\uparrow$ 3}) & 121(\textcolor{teal}{$\downarrow$ 7}) \\
            GPT-3.5 & 132 & 139(\textcolor{red}{$\uparrow$ 7}) & 134(\textcolor{red}{$\uparrow$ 2}) \\
            Yi-Coder-9B & 106 & 137(\textcolor{red}{$\uparrow$ 31}) & 116(\textcolor{red}{$\uparrow$ 10}) \\
            Llama-3.1-8B & 86 & 93(\textcolor{red}{$\uparrow$ 7}) & 67(\textcolor{teal}{$\downarrow$ 19}) \\
            Qwen2.5-Coder-7B & 79 & 79(-) & 84(\textcolor{red}{$\uparrow$ 5}) \\
            Qwen2.5-Coder-3B & 87 & 93(\textcolor{red}{$\uparrow$ 6}) & 92(\textcolor{red}{$\uparrow$ 5}) \\
            Stable-Code-3B & 58 & 60(\textcolor{red}{$\uparrow$ 2}) & 56(\textcolor{teal}{$\downarrow$ 2}) \\
            \bottomrule
        \end{tabular}%
    \label{cot_tot}
\end{table}

%% file: tab/mcts_result.tex
\begin{table}[htbp]
    \centering
    \caption{Correct fix comparison between Vanilla LLMs, CoT and \toolname{} (patch size $\leq$ 32).}
    \begin{tabular}{lcccccc}
        \toprule
        Patch Size & 4 & 8 & 12 & 16 & 32 \\ 
        \midrule
        Qwen2.5-Coder-3B (Vanilla) & 54 & 69 & 80 & 87 & - \\
        Qwen2.5-Coder-3B (CoT)     & 59 & 79 & 88 & 93 & - \\
        Qwen2.5-Coder-3B (\toolname{}) & 58 & 78 & 87 & \textbf{95} & - \\
        \midrule
        Stable-Code-3B (Vanilla)   & 36 & 47 & 54 & 58 & - \\
        Stable-Code-3B (CoT)       & 37 & 49 & 55 & 60 & - \\
        Stable-Code-3B (\toolname{}) & 37 & 50 & 57 & \textbf{62} & - \\
        \midrule
        Qwen2.5-Coder-7B (Vanilla) & 45 & 63 & 69 & 79 & - \\
        Qwen2.5-Coder-7B (CoT)     & 60 & 81 & 94 & 99 & - \\
        Qwen2.5-Coder-7B (\toolname{}) & 65 & 87 & 100 & \textbf{107} & - \\
        \midrule
        Llama-3.1-8B (Vanilla)   & 61 & 74 & 81 & 86 & - \\
        Llama-3.1-8B (CoT)       & 55 & 76 & 88 & 93 & - \\
        Llama-3.1-8B (\toolname{}) & 56 & 72 & 87 & \textbf{97} & - \\
        \midrule
        Yi-Coder-9B (Vanilla)    & 78 & 94 & 101 & 106 & - \\ 
        Yi-Coder-9B (CoT)        & 100 & 119 & 130 & 137 & - \\
        Yi-Coder-9B (\toolname{}) & 109 & 130 & 138 & \textbf{143} & - \\ 
        \midrule
        GPT-4o-mini (Vanilla)      & 105 & 115 & 123 & 128 & - \\ 
        GPT-4o-mini (CoT)          & 101 & 117 & 126 & 131 & - \\
        GPT-4o-mini (\toolname{})  & 127 & 147 & 155 & \textbf{158} & - \\ 
        \midrule
        GPT-3.5 (Vanilla)          & 111 & 120 & 125 & 132 & - \\ 
        GPT-3.5 (CoT)              & 118 & 127 & 132 & 139 & - \\
        GPT-3.5 (\toolname{})      & 134 & 148 & 154 & 159 & \textbf{201}   \\ 
        \bottomrule
    \end{tabular}
    \label{detailed_results}
\end{table}

%% file: tab/patchsize_500_extra_fix.tex
\begin{table}[htp]
    \centering
    \caption{\toolname{} (GPT-3.5) can fix 11 more bugs with larger patch size (32 $\xrightarrow{}$ 500) on Defects4J.}\label{extra_fix_patch_500}
    \begin{tabular}{ll}
        \toprule
        Project & Bugfix \\ 
        \midrule
        Chart & 3 {\color{teal}\ding{51}} \\
        Cli & 25 {\color{teal}\ding{51}}, 14 {\color{red}\ding{55}}, 19 {\color{teal}\ding{51}}, 38 {\color{teal}\ding{51}}\\
        Closure & 53 {\color{red}\ding{55}}, 55  {\color{teal}\ding{51}}, 104 {\color{teal}\ding{51}} \\
        Codec & 2 {\color{teal}\ding{51}}\\
        JacksonDatabind & 17 {\color{teal}\ding{51}}\\
        Jsoup & 26 {\color{teal}\ding{51}}, 55 {\color{teal}\ding{51}}, 75 {\color{red}\ding{55}} \\
         Math & 48 {\color{red}\ding{55}} , 58 {\color{red}\ding{55}} \\
         Time & 15 {\color{teal}\ding{51}} \\
        \bottomrule
    \end{tabular}
\end{table}

%% file: tab/Condefects.tex
\begin{table}[htbp]
    \centering
    \caption{Results on ConDefects (correct/plausible fix).}
    \resizebox{\linewidth}{!}{\begin{tabular}{ccccc}
    
    \toprule
    ChatRepair & GPT-3.5 & AlphaRepair & \toolname{} (48 patch) & \toolname{} (96 patch) \\
    \midrule
        241/249 & 165/171 & 142/160 & 204/211 & 264/287\\
    \bottomrule

    \end{tabular}
    }
    \label{tab:condefects_result}
\end{table}

%% file: tab/swe_result.tex
\begin{table}[t]
\caption{Results on SWE-Bench-Lite.}
\label{tab:swe_result}
\resizebox{1.0\columnwidth}{!}{
\begin{tabular}{llccc}
\toprule
\rowcolor[HTML]{EFEFEF}
\multicolumn{1}{c}{\textbf{SWE System}} & \multicolumn{1}{c}{\textbf{Base Model}} & \textbf{Resolved} & \textbf{\%Resolved} & \textbf{Date} \\
\midrule
\raisebox{-0.15em}{\includegraphics[height=1em]{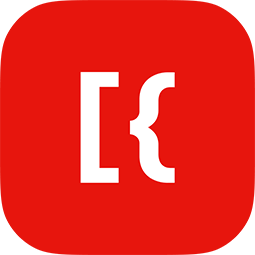}}
Refact.ai Agent  & NA        & 180 & 60\% & 2025-06-25    \\
\raisebox{-0.20em}{\includegraphics[height=1em]{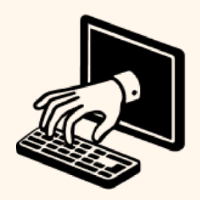}}
SWE-agent~\cite{yang2024swe}                & \raisebox{-0.20em}{\includegraphics[height=1em]{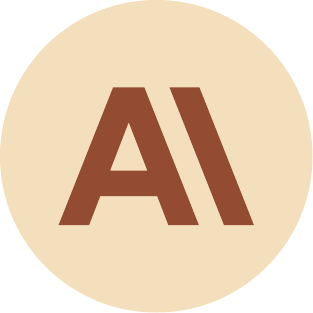}} Claude-4 Sonnet        & 170 & 56.67\% & 2025-05-26    \\
\rowcolor[HTML]{DFDCEF}
\toolname{} (Ours)               &  \raisebox{-0.20em}{\includegraphics[height=1em]{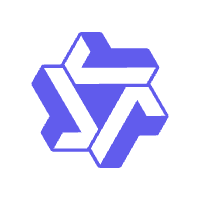}} Qwen3-Coder-480B    & 164 & 54.67\% & 2026-03-22    \\
\raisebox{-0.15em}{\includegraphics[height=1em]{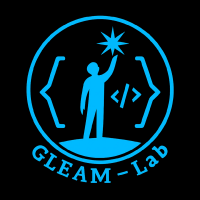}}
KGCompass~\cite{yang2025enhancing} & \raisebox{-0.20em}{\includegraphics[height=1em]{icons/claude.pdf}} Claude-3.5 Sonnet        & 138 & 46\% & 2025-06-19    \\
\rowcolor[HTML]{DFDCEF}
ChatRepair               &  \raisebox{-0.20em}{\includegraphics[height=1em]{icons/qwen.png}} Qwen3-Coder-480B     & 129 & 43\% & 2026-03-22    \\
\raisebox{-0.20em}{\includegraphics[height=1em]{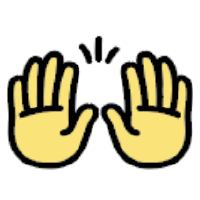}}
OpenHands~\cite{wang2025openhands}           & \raisebox{-0.20em}{\includegraphics[height=1em]{icons/claude.pdf}} Claude-3.5 Sonnet & 125 & 41.67\% & 2024-10-25 \\
\rowcolor[HTML]{DFDCEF}
Vanilla LLMs               &  \raisebox{-0.20em}{\includegraphics[height=1em]{icons/qwen.png}} Qwen3-Coder-480B     & 113 & 37.67\% & 2026-03-22    \\

\bottomrule
\end{tabular}}
\end{table}

%% file: tab/vul4j.tex
\begin{table}[!h]
\centering
\footnotesize
\caption{Results on VUL4J.}
\resizebox{\linewidth}{!}{
\begin{tabular}{c|ccccccc}
\toprule
\textbf{}        & \toolname{}-Vul & FSV-Codex~\cite{effective} & FSV-finetuned~\cite{effective} & NTR~\cite{ntr} & VRPILOT~\cite{vrpilot} & APR4Vul~\cite{bui2024apr4vul} & ChatRepair~\cite{DBLP:conf/issta/Xia024}  \\ 
\midrule
CF   & 27/79    & 10.9/79      & 9/79             & 14/79  & 14/79 & 16/79 & 15/79    \\
RCR & 34.17\% & 13.79\% & 11.39\% & 17.72\% & 17.72\% & 20.25\% & 18.98\% \\
\bottomrule 
\end{tabular}}
\label{tab:vul}
\end{table}

%% file: tab/cost.tex
\begin{table}[htbp]
    \centering
    \caption{Cost analysis on Defects4J.}
    \resizebox{\linewidth}{!}{\begin{tabular}{lccccc}
    \toprule
    Method & Patch/Bug & Time/Bug & Token/Bug & Money/Bug & Charge/1k tokens\\
    \midrule
        ChatRepair (2024)~\cite{sobania2023analysisautomaticbugfixing}& 500 & \revise{$\leq$} 5 hours & 210,000 & \$0.42 & \$0.002\\
        ChatRepair (today's price) & 500 & \revise{$\leq$} 5 hours & 210,000 & \$0.14 & -\\
        RepairAgent (2024)~\cite{bouzenia2024repairagentautonomousllmbasedagent} & 117  & 920 seconds & 270,000 & \$0.14 & - \\
        \toolname{} (2026) & 16 & 23.64 min & 20,000 & \$0.03 & \$0.0015\\
        \toolname{} (2026) & 32& 50 min & 40,000 & \$0.06 & \$0.0015\\
    \bottomrule
    \end{tabular}
    }
    \label{tab:cost_analysis}
\end{table}